\documentclass[10pt, a4paper]{article}
\usepackage{lrec}
\usepackage{flushend}

\title{CRWIZ: A Framework for Crowdsourcing Real-Time Wizard-of-Oz Dialogues}

\name{Francisco J. Chiyah Garcia, José Lopes, Xingkun Liu, Helen Hastie}

\address{School of Mathematical and Computer Sciences, \\Heriot-Watt University, \\
         Edinburgh, United Kingdom, \\
         \{fjc3, jd.lopes, x.liu, h.hastie\}@hw.ac.uk\\}

\abstract{
Large corpora of task-based and open-domain conversational dialogues are hugely valuable in the field of data-driven dialogue systems. Crowdsourcing platforms, such as Amazon Mechanical Turk, have been an effective method for collecting such large amounts of data. However, difficulties arise when task-based dialogues require expert domain knowledge or rapid access to domain-relevant information, such as databases for tourism. This will become even more prevalent as dialogue systems become increasingly ambitious, expanding into tasks with high levels of complexity that require  collaboration and forward planning, such as in our domain of emergency response. In this paper, we propose CRWIZ: a framework for collecting real-time Wizard of Oz dialogues through crowdsourcing for collaborative, complex tasks. This framework uses semi-guided dialogue to avoid interactions that breach procedures and processes only known to experts, while enabling the capture of a wide variety of interactions. The framework is available at 
\href{https://github.com/JChiyah/crwiz}{\texttt{https://github.com/JChiyah/crwiz}}.
\\ \newline  
}

\begin{document}

\maketitleabstract

\section{Introduction}

Recent machine learning breakthroughs in dialogue systems and their respective components have been made possible by training on publicly available large scale datasets, such as ConvAI \cite{Logacheva2018}, bAbI \cite{WestonBCM15bAbI} and MultiWoZ \cite{budzianowski2018multiwoz}, many of which are collected on crowdsourcing services, such as Amazon Mechanical Turk and Figure-eight.  These data collection methods have the benefits of being cost-effective, time-efficient to collect and scalable,  enabling the collection of large numbers of dialogues.  

Where this crowdsourcing method has its limitations is when specific domain expert knowledge is required, rather than general conversation. These tasks include, for example, call centre agents  \cite{peskov2019multi} or clerks with access to a database, as is required for tourism information and booking \cite{budzianowski2018multiwoz}. In the near future, there will be a demand to extend this to workplace-specific tasks and procedures. Therefore, a method of gathering crowdsourced dialogue data is needed that ensures compliance with such procedures, whilst providing coverage of a wide variety of dialogue phenomena that could be observed in deployment of a trained dialogue system. 

Wizard-of-Oz data collections in the past have provided such a mechanism. However, these have traditionally not been scalable because of the scarcity of Wizard experts or the expense to train up workers.  This was the situation with an initial study reported in \cite{HRILopes2019}, which was conducted in a traditional lab setting and where the Wizard (an academic researcher) had to learn, through training and reading manuals, how best to perform operations in our domain of emergency response. 

We present the CRWIZ Intelligent Wizard Interface that enables a crowdsourced Wizard to make intelligent, relevant choices without such intensive training by providing a restricted list of valid and relevant dialogue task actions, which changes dynamically based on the context, as the interaction evolves.

Prior crowdsourced wizarded data collections have divided the dialogue up into turns and each worker's job consists of one turn utterance generation given a static dialogue context, as in the MultiWoZ dataset \cite{budzianowski2018multiwoz}. However, this can limit naturalness of the dialogues by restricting forward planning, collaboration and use of memory that humans use for complex multi-stage tasks in a shared dynamic environment/context.

Our scenario is such a complex task. Specifically, our scenario relates to using robotics and autonomous systems on an offshore energy platform to resolve an emergency and is part of the EPSRC ORCA Hub project \cite{ORCA}. The ORCA Hub vision is to use teams of robots and autonomous intelligent systems  to work on offshore energy platforms to enable cheaper, safer and more efficient working practices. An important part of this is ensuring safety of robots in complex, dynamic and cluttered environments, co-operating with remote operators. With this data collection method reported here, we aim to automate a conversational Intelligent Assistant (Fred), who acts as an intermediary between the operator and the multiple robotic systems \cite{ChiyahHRI20Workshop,LopesHRI20Demo}. Emergency response is clearly a high-stakes situation, which is difficult to emulate in a lab or crowdsourced data collection environment. Therefore, in order to foster engagement and collaboration, the scenario was gamified with a  monetary reward given for task success. 





In this paper, we provide a brief survey of existing datasets and describe the CRWIZ framework for pairing crowdworkers and having half of them acting as Wizards by limiting their dialogue options only to relevant and plausible ones, at any one point in the interaction. We then perform a data collection and compare our dataset to a similar dataset collected in a more controlled lab setting with a single Wizard \cite{HRILopes2019} and discuss the advantages/disadvantages of both approaches. Finally, we present future work. 
Our contributions are as follows:
\\
\begin{itemize}
\item The release of a platform for the CRWIZ Intelligent Wizard Interface to allow for the collection of dialogue data for longer complex tasks, by providing a dynamic  selection of relevant dialogue acts.  
\item A survey of existing datasets and data collection platforms, with a comparison to the CRWIZ data collection for Wizarded crowdsourced data 
in task-based interactions.
\end{itemize}

\begin{table*}[!ht]
\begin{center}
\begin{tabularx}{\textwidth}{|m{5.65cm}|m{1.6cm}|m{1.7cm}|l|m{1.6cm}|m{1.65cm}|}
    \hline
    \textbf{Dataset} & \textbf{WoZ?} & \textbf{Single Participant?} & \textbf{Crowdsourced?} & \textbf{Interaction Modality} & \textbf{Domain} \\
    \hline
    MultiWoZ \cite{budzianowski2018multiwoz} & Partially$\dagger$ & No & Yes & Text & Tourism \\
    \hline
    RDG-Image Game \cite{Manuvinakurike2015} & No & Yes & Yes & Speech & Image game \\
    \hline
    MultiDiaGo \cite{peskov2019multi} & Controlled Wizards & N/A & User only & Text & Fast food, airline, finance, etc. \\
    \hline
    Stanford Multi-Domain Dialog Data \cite{Eric2017Key} & Partially$\dagger$ & No & Yes & Text & Car assistant \\
    \hline
    Cambridge Restaurant \cite{wen-etal-2017-network} & Partially$\dagger$ & No & Yes & Text & Restaurants \\
    \hline
    MetalWoz \cite{lee2019multi-domain} & N/A & Yes & Yes & Text & Multiple domains \\
    \hline
    AirDialogue \cite{wei-etal-2018-airdialogue} & N/A & N/A & Yes & Text & Flight booking \\
    \hline
    TalkTheWalk \cite{de2018talk} & No & No & Yes & Text/images & Navigation \\
    \hline
    Deal or No Deal \cite{lewis-etal-2017-deal} & No & Yes & Yes & Text & Negotiation \\
    \hline
    Frames \cite{ElAsri2017} & Partially$\dagger$ & Yes & No & Text & Tourism \\
    \hline
    ConvAI \cite{Logacheva2018} & No & Yes & Yes$\ddagger$ & Text & Context-based chat \\
    \hline
    bAbI Dialogues \cite{WestonBCM15bAbI} & No & Artificial data & No & Text & Restaurants \\
    \hline
    EDINA \cite{DBLP:journals/corr/abs-1709-09816} & No & Yes & Yes & Text & Chitchat \\
    \hline
    Fantom \cite{jonell2019crowdsourcing} & No & No & Yes & Text and speech & Chitchat \\
    \hline
    MutualFriends \cite{he-etal-2017-learning} & No & Yes & Yes & Text & Context-based chat \\
    \hline
    Taskmaster-1 \cite{Byrne2019} & Controlled Wizards & Yes & User only & Text and speech & Multiple domains \\
    \hline
    Collaborative Planning Corpus \cite{katsakioris-etal-2019-corpus} & Yes & Yes & No & Text and images & Mission planning \\
    \hline
    \textbf{Our Data} & \textbf{Yes} & \textbf{Yes} & \textbf{Yes} & \textbf{Text} & \textbf{Emergency response\footnotemark} \\
    \hline
\end{tabularx}
\caption{Comparison of relevant recent works. In order, the columns refer to: the dataset and reference; if the dataset was generated using Wizard-of-Oz techniques; if there was a unique participant per role for the whole dialogue; if the dataset was crowdsourced; the type of interaction modality used; and finally, the type of task or domain that the dataset covers. 
$\dagger$ The participants were aware that the dialogue was authored by humans.
$\ddagger$ The participants were volunteers without getting paid.
}
\label{tab:work_comparison}
\end{center}
\end{table*}

\section{Related Work}

Table \ref{tab:work_comparison} gives an overview of prior work and datasets. We report various factors to compare to the CRWIZ dataset corresponding to columns in Table \ref{tab:work_comparison}: whether or not the person was aware they were talking to a bot; 
whether each dialogue had a single or multiple participants per role; whether the data collection was crowdsourced; and the modality of the interaction and the domain. As we see from the bottom row, none of the datasets reported in the table meet all the criteria we are aiming for, exemplifying the need for a new and novel approach.



Collecting large amounts of dialogue data can be very challenging as two interlocutors are required to create a conversation. If one of the partners in the conversation is a machine as in \cite{Logacheva2018}, the challenge becomes slightly easier since only one partner is lacking. However, in most cases these datasets are aimed at creating resources to train the  conversational system itself. Self-authoring the dialogues \cite{DBLP:journals/corr/abs-1709-09816} or artificially creating data \cite{WestonBCM15bAbI} could be a solution to rapidly collect data, but this solution has been shown to produce low quality unnatural data \cite{jonell2019crowdsourcing}. 

One way to mitigate the necessity of pairing two users simultaneously is to allow several participants to contribute to the dialogue, one turn at the time. This approach has been used both in task-oriented \cite{wen-etal-2017-network,budzianowski2018multiwoz,Eric2017Key} and chitchat \cite{jonell2019crowdsourcing}. This means that the same dialogue can be authored by several participants. However, this raises issues in terms of coherence and forward-planning. These can be addressed by carefully designing the data collection to provide the maximum amount of information to the participants (e.g. providing the task, personality traits of the bot, goals, etc.) but then this adds to cognitive load, time, cost and participant fatigue.

Pairing is a valid option, which has been used in a number of recent data collections in various domains, such as navigating in a city \cite{de2018talk}, playing a negotiation game \cite{lewis-etal-2017-deal}, talking about a person \cite{he-etal-2017-learning}, playing an image game \cite{Manuvinakurike2015} or having a chat about a particular image that is shown to both participants \cite{Ilinykh2019,das2017visual}. Pairing frameworks exist such as \texttt{Slurk} \cite{schlangen2018slurk}. Besides its pairing management feature, \texttt{Slurk} is designed in order to allow researchers to modify it and implement their own data collection rapidly. 

The scenarios for the above-mentioned data collections are mostly intuitive tasks that humans do quite regularly, unlike our use-case scenario of emergency response. Role playing is one option. For example, recent work has tried to create datasets for non-collaborative scenarios \cite{li2019endtoend,wang-etal-2019-persuasion}, requesting participants to incarnate a particular role during the data collection. This is particularly challenging when the recruitment is done via a crowdsourcing platform. In \cite{wang-etal-2019-persuasion}, the motivation for the workers to play the role is intrinsic to the scenario. In this data collection, one of the participants tries to persuade their partner to contribute to a charity with a certain amount of money. As a result of their dialogue, the money that the persuadee committed to donate was actually donated to a charity organising.
However, for scenarios such as ours, the role playing requires a certain expertise and it is questionable whether the desired behaviour would be achieved simply by letting two non-experts converse with free text.

Therefore, in recent data collections, there have been a number of attempts to control the data quality in order to produce a desired behaviour. For example, in \cite{ElAsri2017}, the data collection was done with a limited number of subjects who performed the task several days in a row, behaving both as the Wizard and the customer of a travel agency. The same idea was followed in \cite{wei-etal-2018-airdialogue}, where a number of participants took part in the data collection over a period of 6 months and, in \cite{peskov2019multi,Byrne2019} where a limited number of subjects were trained to be the Wizard. This quality control, however, naturally comes with the cost of recruiting and paying these subjects accordingly.

The solution we propose in this paper tries to minimise these costs by increasing the pool of Wizards to anyone wanting to collaborate in the data collection, by providing them the necessary guidance to generate the desired dialogue behaviour. This is a valuable solution for collecting dialogues in domains where specific expertise is required and the cost of training capable Wizards is high. We required fine-grained control over the Wizard interface so as to be able to generate more directed dialogues for specialised domains, such as emergency response for offshore facilities. By providing the Wizard with several dialogue options (aside from free text), we guided the conversation and could introduce actions that change an internal system state.  This proposes several advantages:

\begin{enumerate}
    \item A guided dialogue allows for set procedures to be learned and reduces the amount of data needed for a machine learning model for dialogue management to converge.
    \item Providing several dialogue options to the Wizard increases the pace of the interaction and allows them to understand and navigate more complex scenarios.
\end{enumerate}

\section{System Overview}

\begin{figure}[t!]
  \centering
  \includegraphics[width=1\columnwidth]{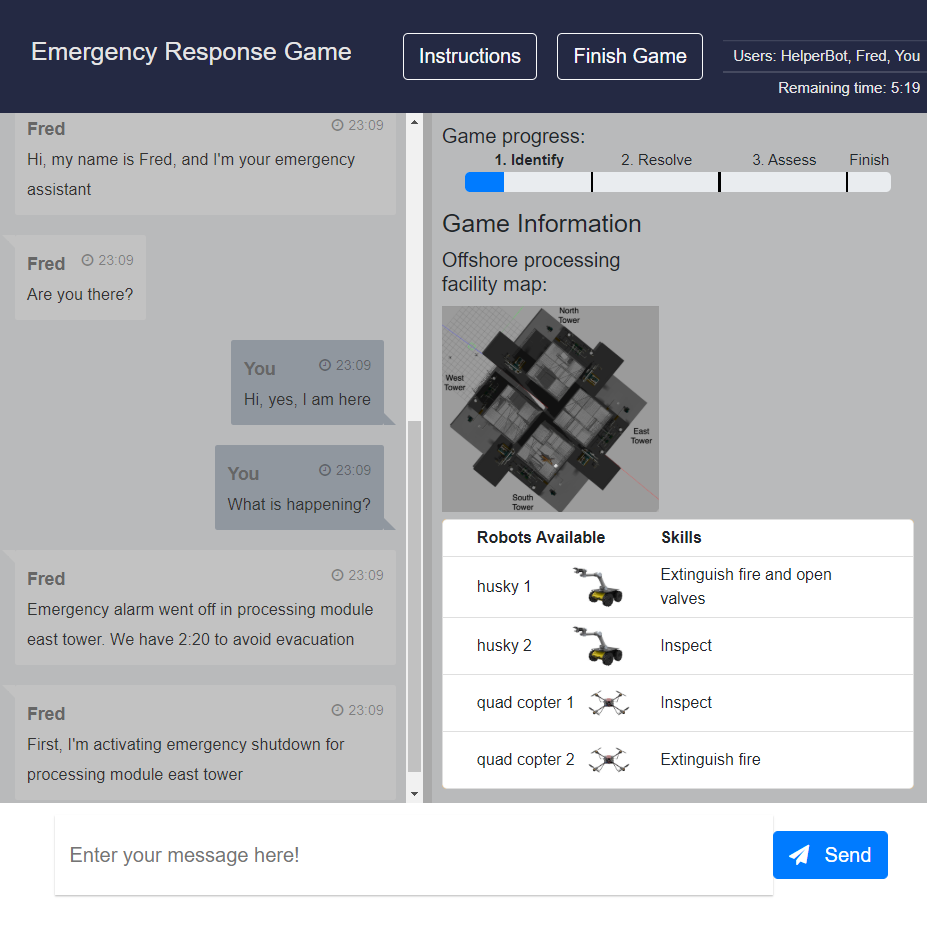}
  \caption{Interface shown to those in the Operator role running on the \texttt{Slurk} interaction server. It has a similar layout to other chat applications with the chat window on the left and a field to send messages at the bottom. The right side is used to display additional information.}
  \label{fig:interface_operator}
\end{figure}

The CRWIZ Intelligent Wizard Interface resides on \texttt{Slurk} \cite{schlangen2018slurk}, an interaction server built for conducting dialogue experiments and data collections. \texttt{Slurk} handles the pairing of participants and provides a basic chat layout amongst other features. Refer to \cite{schlangen2018slurk} for more information on the pairing of participants and the original chat layout. Our chat layout remains similar to \texttt{Slurk} with an important difference. In our scenario, we assign each new participant a role (Operator or Wizard) and, depending on this role, the participant sees different game instructions and chat layout schemes. These are illustrated in Figures \ref{fig:interface_operator} and \ref{fig:interface_wizard}, for the Operator and Wizard respectively. The main components are described in turn below:  1) The Intelligent Wizard Interface; 2) dialogue structure; and 3) system-changing actions.


\textbf{Wizard interface:} the interface shown to participants with the Wizard role provides possible actions on the right-hand side of the browser window. These actions could be verbal, such as sending a message, or non-verbal, such as switching on/off a button to activate a robot.
Figure \ref{fig:interface_wizard} shows this interface with several actions available to be used in our data collection.

\textbf{Dialogue structure:} we introduced structured dialogues through a Finite State Machine (FSM) that controls the current dialogue state and offers multiple suitable and relevant state transitions (actions) to the Wizard depending on the point in the interaction, the state of the world and the history. A graph of dialogue states, transitions and utterances is loaded when the system is initialised, and each chat room has its own dialogue state, which changes through actions. 

\footnotetext{The CRWIZ framework is domain-agnostic, but the data collected with it corresponds to the emergency response domain.}

\textbf{System-changing actions:} actions trigger transitions between the states in the FSM. We differentiate two types of actions:
\begin{enumerate}
    \item Verbal actions, such as the dialogue options available at that moment. The Wizard can select one of several predefined messages to send, or type their own message if needed. Free text messages do not change the dialogue state in the FSM, so it is important to minimise their use by providing enough dialogue options to the Wizard.
    Predefined messages can also trigger other associated events such as pop-ups or follow-up non-verbal actions.
    \item Non-verbal actions, such as commands to trigger events. These can take any form, but we used buttons to control robots in our data collection.
\end{enumerate}

Submitting an action would change the dialogue state in the FSM, altering the set of actions available in the subsequent turn visible to the Wizard. Some dialogue options are only possible at certain states, in a similar way as to how non-verbal actions are enabled or disabled depending on the state. This is reflected in the Wizard interface.



The advantage of the CRWIZ framework is that it can easily be adapted to different domains and procedures
by simply modifying the dialogue states loaded at initialisation. These files are in YAML format and have a simple structure that defines their NLG templates (the FSM will pick one template at random if there is more than one) and the states that it can transition to. Note, that some further modifications may be necessary if the scenario is a slot-filling dialogue requiring specific information at various stages.

Once the dialogue between the participants finishes, they receive a code in the chat, which can then be submitted to the crowdsourcing platform for payment. The CRWIZ framework generates a JSON file in its log folder with all the information regarding the dialogue, including messages sent, FSM transitions, world state at each action, etc. Automatic evaluation metrics and annotations are also appended such as number of turns per participant, time taken or if one of the participants disconnected. Paying the crowdworkers can be done by just checking that there is a dialogue file with the token that they entered.

\section{Data Collection}

\begin{figure}[t!]
  \centering
  \includegraphics[width=1\columnwidth]{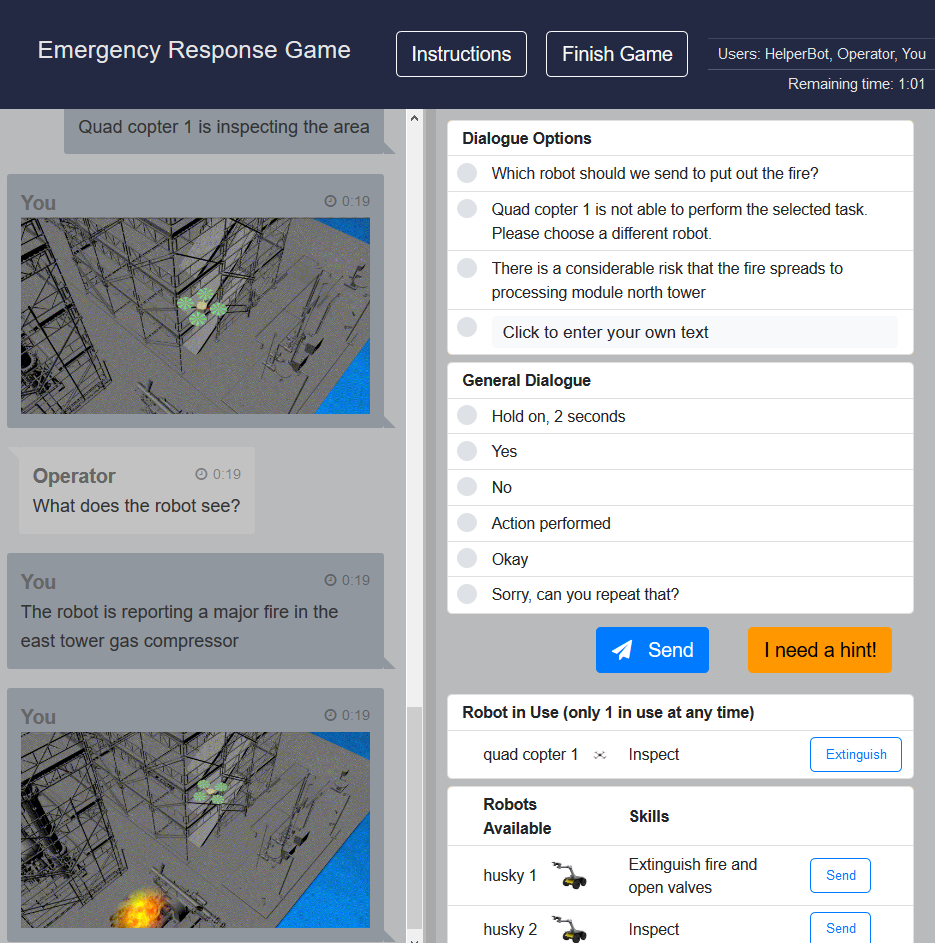}
  \caption{Interface shown to those in the Emergency Assistant Wizard role running on the \texttt{Slurk} interaction server. The chat window is on the left, with the dialogue options and buttons to control the robots on the right. The chat here shows GIFs that appear to increase engagement and show game progress visually.}
  \label{fig:interface_wizard}
\end{figure}

We set up a crowdsourced data collection through Amazon Mechanical Turk, in which two participants chatted with each other in a setting involving an emergency at an offshore facility. As mentioned above, participants had different roles during the interaction: one of them was an Operator of the offshore facility whereas the other one acted as an Intelligent Emergency Assistant. Both of them had the same goal of resolving the emergency and avoiding evacuation at all costs, but they had different functions in the task:

\begin{itemize}
    \item The \textbf{Operator} was responsible for the facility and had to give instructions to the Emergency Assistant to perform certain actions, such as deploying emergency robots. Participants in the role of Operator were able to chat freely with no restrictions and were additionally given a map of the facility and a list of available robots (see Figure \ref{fig:interface_operator}).
    \item The \textbf{Emergency Assistant} had to help the Operator handle the emergency by providing guidance and executing actions. Participants in the role of Emergency Assistant had predefined messages depending on the task progress. They had to choose between one of the options available, depending on which made sense at the time, but they also had the option to write their own message if necessary. The Emergency Assistant role mimics that of the Wizard in a Wizard-of-Oz experiment (see Figure \ref{fig:interface_wizard}).
\end{itemize}

The participants had a limited time of 6 minutes to resolve the emergency, which consisted of the following sub-tasks: 1) identify and locate the emergency; 2) resolve the emergency; and 3) assess the damage caused.
They had four robots available to use with different capabilities: two ground robots with wheels (Husky) and two Quadcopter UAVs (Unmanned Aerial Vehicles). For images of these robots, see Figure \ref{fig:interface_operator}. Some robots could inspect areas whereas others were capable of activating hoses, sprinklers or opening valves. Both participants, regardless of their role, had a list with the robots available and their capabilities, but only the Emergency Assistant could control them. This control was through high-level actions (e.g. moving a robot to an area, or ordering the robot to inspect it) that the Emergency Assistant had available as buttons in their interface, as shown in Figure \ref{fig:interface_wizard}. 
For safety reasons that might occur in the real world, only one robot could be active doing an action at any time. The combinations of robots and capabilities meant that there was not a robot that could do all three steps of the task mentioned earlier (inspect, resolve and assess damage), but the robots could be used in any order allowing for a variety of ways to resolve the emergency. 

\begin{figure}[ht]
  \centering
  \includegraphics[width=0.97\columnwidth]{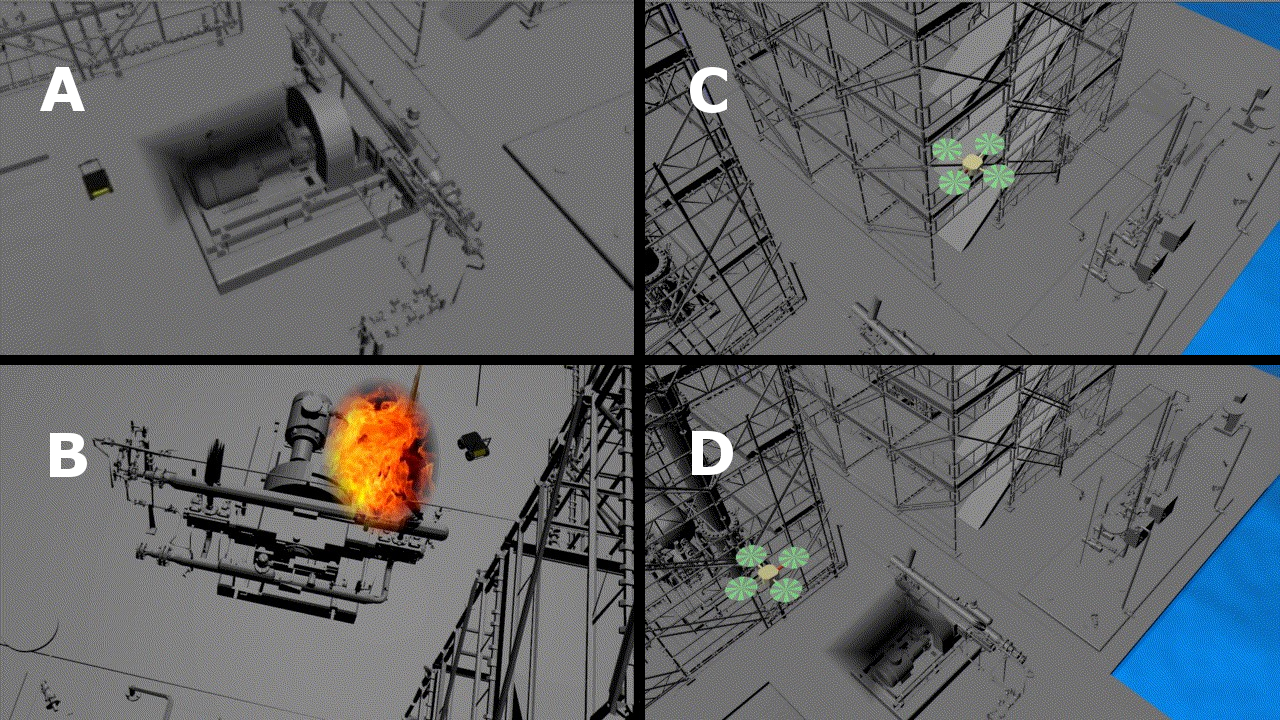}
  \caption{Some of the GIFs shown during the game. A and B are Husky robots assessing damages and inspecting a fire respectively. C and D show Quadcopter UAVs moving and inspecting an area.}
  \label{fig:robot_gif}
\end{figure}

Participants would progress through the task when certain events were triggered by the Emergency Assistant. For instance, inspecting the area affected by an alarm would trigger the detection of the emergency. After locating the emergency, other dialogue options and commands would open up for the Emergency Assistant. In order to give importance to the milestones in the dialogue, these events were also signalled by GIFs (short animated video snippets) in the chat that both participants could see (e.g. a robot finding a fire), as in Figure \ref{fig:robot_gif}. The GIFs were added for several reasons: to increase participant engagement and situation awareness, to aid in the game and to show progress visually. Note that there was no visual stimuli in the original WoZ study \cite{HRILopes2019} but they were deemed necessary here to help the remote participants contextualise the scenario. These GIFs were produced using a Digital Twin simulation of the offshore facility with the various types of robots. See \cite{digitaltwin19} for details on the Digital Twin.

\subsection{Implementation}

The dialogue structure for the Emergency Assistant (the Wizard) followed a dialogue flow previously used for the original lab-based Wizard-of-Oz study \cite{HRILopes2019}
but which was slightly modified and simplified for this crowdsourced data collection. In addition to the transitions that the FSM provides, there are other fixed dialogue options always available such as \textit{``Hold on, 2 seconds''}, \textit{``Okay''} or \textit{``Sorry, can you repeat that?''} as a shortcut for commonly used dialogue acts, as well as the option to type a message freely.

The dialogue has several paths to reach the same states with varying levels of Operator control or engagement that enriched the heterogeneity of conversations. The Emergency Assistant dialogue options show various speaking styles, with a more assertive tone (\textit{``I am sending Husky 1 to east tower''}) or others with more collaborative connotations (\textit{``Which robot do you want to send?''} or \textit{``Husky 1 is available to send to east tower''}). Refer to \cite{HRILopes2019} for more details. Furthermore, neither participants were restricted in the number of messages that they could send and we did not require a balanced number of turns between them. However, there were several dialogue transitions that required an answer or authorisation from the Operator, so the FSM would lock the dialogue state until the condition was met.
As mentioned earlier, the commands to control the robots are also transitions of the FSM, so they were not always available.

The Emergency Assistant interface contains a button to get a hint if they get stuck at any point of the conversation. This hint mechanism, when activated, highlights one of the possible dialogue options or robot buttons. This highlighted transition was based on the observed probability distribution of transitions from \cite{HRILopes2019} to encourage more collaborative interaction than a single straight answer.

As in the real world, robot actions during the task were simulated to take a certain period of time, depending on the robot executing it and the action. The Emergency Assistant had the option to give status updates and progress reports during this period. Several dialogue options were available for the Emergency Assistant whilst waiting. The time that robots would take to perform actions was based on simulations run on a Digital Twin of the offshore facility implemented in Gazebo \cite{digitaltwin19}. 
Specifically, we pre-simulated typical robot actions, with the robot's progress and position reflected in the Wizard interface with up-to-date dialogue options for the Emergency Assistant. Once the robot signals the end of their action, additional updated dialogue options and actions are available for the Emergency Assistant. This simulation allowed us to collect dialogues with a realistic embedded world state.

\subsection{Deployment}

We used Amazon Mechanical Turk (AMT) for the data collection. We framed the task as a game to encourage engagement and interaction. The whole task, (a Human Intelligence Task (HIT) in AMT) consisted of the following:

\begin{enumerate}
    \item Reading an initial brief set of instructions for the overall task.
    \item Waiting for a partner for a few seconds before being able to start the dialogue.
    \item When a partner was found, they were shown the instructions for their assigned role. As these were different, we ensured that they both took around the same time. The instructions had both a text component and a video explaining how to play, select dialogues, robots, etc\footnote{Video with instructions for the emergency assistant is available at \href{http://bit.ly/32Rjg8N}{http://bit.ly/32Rjg8N}}.
    \item Playing the game to resolve the emergency. This part was limited to 6 minutes.
    \item Filling a post-task questionnaire about partner collaboration and task ease.
\end{enumerate}

The participants received a game token after finishing the game that would allow them to complete the questionnaire and submit the task. This token helped us link their dialogue to the responses from the questionnaire. 


Several initial pilots helped to define the total time required as 10 minutes for all the steps above. We set the HIT in AMT to last 20 minutes to allow additional time should any issues arise.  The pilots also helped setting the payment for the workers. Initially, participants were paid a flat amount of \$1.4 per dialogue. However, we found that offering a tiered payment tied to the length of the dialogue and bonus for completing the task was the most successful and cost-effective method to foster engagement and conversation: 

\begin{itemize}
    \item \$0.5 as base for attempting the HIT, reading the instructions and completing the questionnaire. 
    \item \$0.15 per minute during the game, for a maximum of \$0.9 for the 6 minutes.
    \item \$0.2 additional bonus if the participants were able to successfully avoid the evacuation of the offshore facility.
\end{itemize}

The pay per worker was therefore \$1.4 for completing a whole dialogue and \$1.6 for those who resolved the emergency for a 10-minute HIT. This pay is above the Federal minimum wage in the US (\$7.25/hr or ~\$0.12/min) at the time of the experiment.

The post-task questionnaire had four questions rated in 7-point rating scales that are loosely based on the PARADISE \cite{Walker1997} questions for spoken dialogue systems:

\begin{enumerate}
    \item[Q1.] \textbf{Partner collaboration:} \textit{``How helpful was your partner?''} on a scale of 1 (not helpful at all) to 7 (very helpful).
    \item[Q2.] \textbf{Information ease:} \textit{``In this conversation, was it easy to get the information that I needed?''} on a scale of 1 (no, not at all) to 7 (yes, completely).
    \item[Q3.] \textbf{Task ease:} \textit{``How easy was the task?''} on a scale of 1 (very easy) to 7 (very difficult).
    \item[Q4.] \textbf{User expertise:} \textit{``In this conversation, did you know what you could say or do at each point of the dialog?''} on a scale of 1 (no, not at all) to 7 (yes, completely).
\end{enumerate}

At the end, there was also an optional entry to give free text feedback about the task and/or their partner.

\begin{table*}[!ht]
\begin{center}
\begin{tabular}{|l|c|c|}
    \hline
    & \textbf{Dialogues Collected} 
    & \textbf{Lopes et al. (2019)} \\
    \textbf{Feature} & \textbf{Mean (SD)} & \textbf{Mean (SD)}  \\
    \hline
    Number of Turns & 25.22 (9.69) & 53.26 (9.13) \\
    \hline
    Number of Operator Turns & 7.99 (3.96) & 9.78 (7.67) \\
    \hline
    Number of Emergency Assistant Turns & 17.23 (7.97) & 43.64 (4.45) \\
    \hline
    Operator Turn Length (words) & 3.88 (1.69) & 3.02 (1.59) \\
    \hline
    Emergency Assistant \% typed Utterances & 2.29\% (5.16\%) & 1.72\% (3.34 \%) \\
    \hline
\end{tabular}
\caption{Interaction features of the dialogues collected. 
We compare it with the results of the Wizard-of-Oz experiment in a controlled setting from \cite{HRILopes2019}.}
\label{tab:interaction_stats}
\end{center}
\end{table*}

\begin{figure}[t]
  \centering
  \includegraphics[width=0.95\columnwidth]{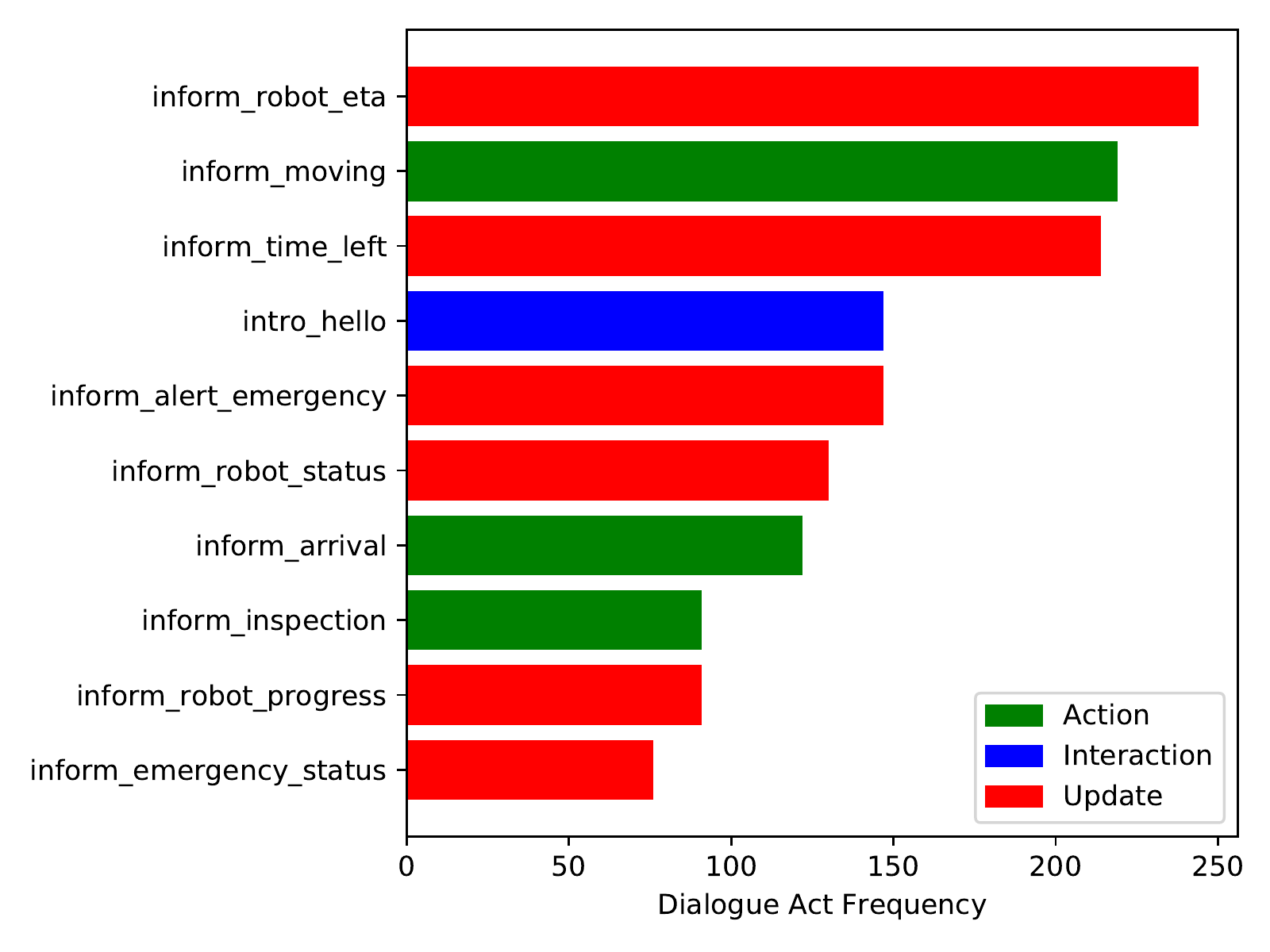}
  \caption{Frequency of the top-10 Emergency Assistant dialogue acts in the data collected. There were 40 unique dialogue acts, each with two or more distinct formulations on average. Most of them also had slots to fill with contextual information, such as the name of the robot. Dialogue acts are colour-coded based on 3 main types.} 
  \label{fig:dialogue_acts_amt}
\end{figure}

\begin{figure}[t]
  \centering
  \includegraphics[width=0.95\columnwidth]{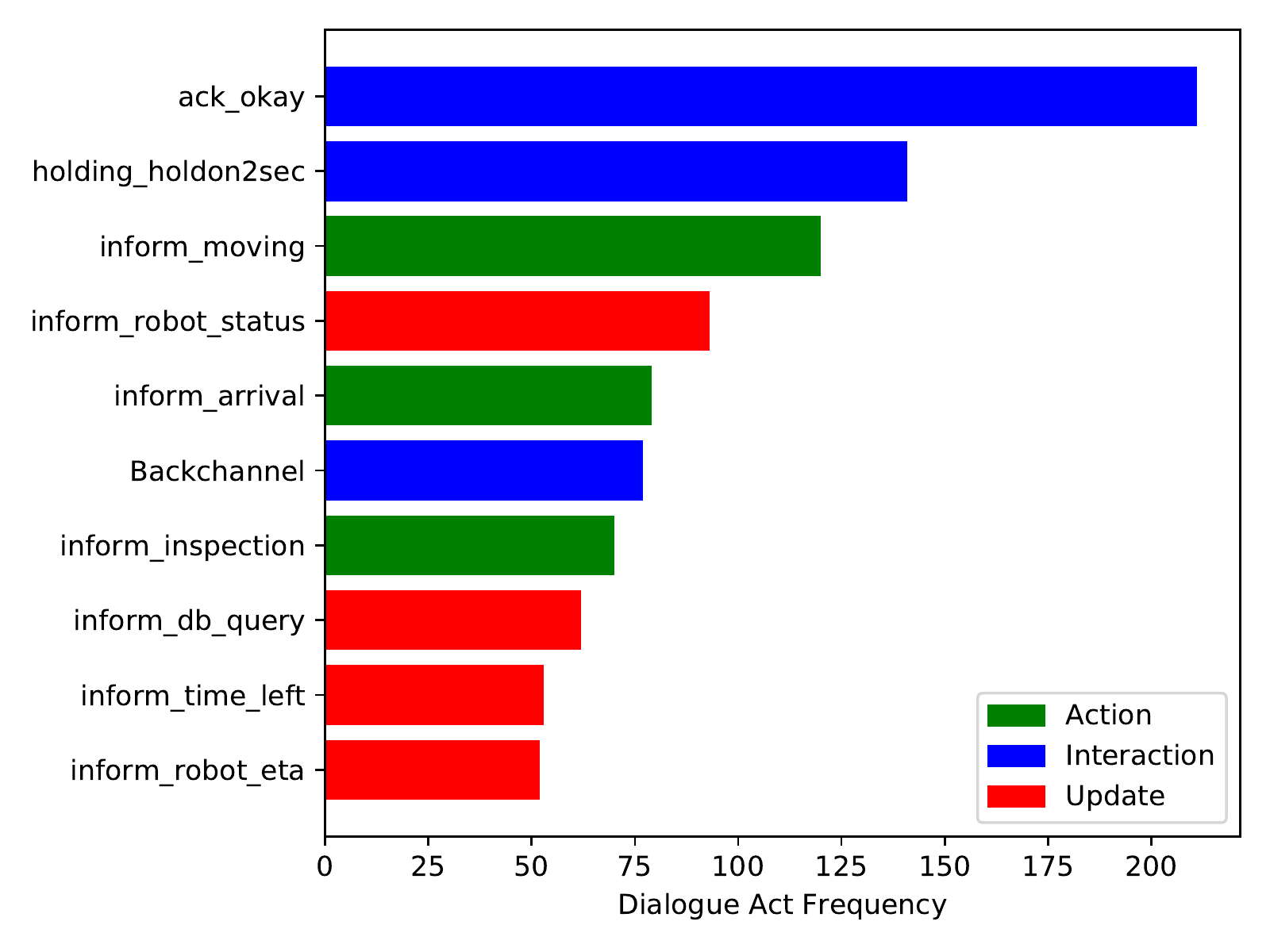}
  \caption{Frequency of the top-10 Emergency Assistant dialogue acts in \cite{HRILopes2019}.}
  \label{fig:dialogue_acts_hri}
\end{figure}

\section{Data Analysis}




For the intitial data collection using the CRWIZ platform, 145 unique dialogues were collected (each dialogue consists of a conversation between two participants). All the dialogues were manually checked by one of the authors and those where the workers were clearly not partaking in the task or collaborating were removed from the dataset.  The average time per assignment was 10 minutes 47 seconds, very close to our initial estimate of 10 minutes, and the task was available for 5 days in AMT.
Out of the 145 dialogues, 14 (9.66\%) obtained the bonus of \$0.2 for resolving the emergency. 
We predicted that only a small portion of the participants would be able to resolve the emergency in less than 6 minutes, thus it was framed as a bonus challenge rather than a requirement to get paid\footnote{Dialogues where the emergency was not resolved are still valid.}. The fastest time recorded to resolve the emergency was 4 minutes 13 seconds with a mean of 5 minutes 8 seconds. Table \ref{tab:interaction_stats} shows several interaction statistics for the data collected compared to the single lab-based WoZ study \cite{HRILopes2019}.





\begin{table}[ht]
\footnotesize
    \centering
    \begin{tabular}{|l|c|c|}
    \hline
         \textbf{Type of DA} & \textbf{Dialogues Collected} & \textbf{\cite{HRILopes2019}} \\
         \hline
         \% Request &  7.14 & 6.85 \\
         \hline
         \% Interaction & 20.31 & 29.20 \\
         \hline
         \% Action & 20.19 & 21.40 \\
         \hline
         \% Update & 52.36 & 42.54 \\
         \hline
    \end{tabular}
    \caption{Distribution of the types of dialogue acts in the data collected with CRWIZ, compared with \cite{HRILopes2019}.}
    \label{tab:da_type_distribution}
\end{table}

\newcolumntype{Y}{>{\centering\arraybackslash}X}

\begin{table*}[!ht]
\begin{center}
\begin{tabularx}{\textwidth}{|l|c|Y|Y|}
    \hline
     & \textbf{Dialogues Collected (145)} & \textbf{Emergency Not Resolved Dialogues (131)} & \textbf{Emergency Resolved Dialogues (14)} \\
     & Mean/Median/Mode (SD) & Mean/Median/Mode (SD) & Mean/Median/Mode (SD) \\
    \hline
    Q1. Partner collaboration & 3.76/4/1 (2.0) & 3.59/4/1 (1.99) & \textbf{5.19/5/5} (1.52)* \\
    \hline
    Q2. Information ease & 3.65/4/1 (2.0) & 3.55/3/1 (2.01) & \textbf{4.48/5/5} (1.72)* \\
    \hline
    Q3. Task ease & 3.08/3/2 (1.73) & 3.03/3/2 (1.74) & \textbf{3.56/3/3} (1.67) \\
    \hline
    Q4. User expertise & 4.09/4/4 (1.81) & 4.03/4/4 (1.82) & \textbf{4.59/5/6} (1.67) \\
    \hline
\end{tabularx}
\caption{Subjective ratings for the post-task survey reporting Mean, Median, Mode and Standard Deviation (SD). Scales were on a 7-point rating scale. ``Dialogues Collected'' refers to all the dialogues collected after filtering, whereas the other columns are for the dialogues that did not resolved the emergency (``Emergency Not Resolved Dialogues'') and those that did (``Emergency Resolved Dialogues''). Higher is better (Q3 reversed for this table). Highest numbers are \textbf{bold}. * indicates significant differences ($p < 0.05$, Mann-Whitney-U) between Emergency Resolved and Emergency Not Resolved dialogues.} 
\label{tab:descriptive_stats}
\end{center}
\end{table*}

\paragraph{Subjective Data}
Table \ref{tab:descriptive_stats} gives the results from the post-task survey. We observe, that subjective and objective task success are similar in that the dialogues that resolved the emergency were rated consistently higher than the rest.


Mann-Whitney-U one-tailed tests show that the scores of the Emergency Resolved Dialogues for Q1 and Q2 were significantly higher than the scores of the Emergency Not Resolved Dialogues at the 95\% confidence level (Q1: $U = 1654.5$, $p < 0.0001$; Q2: $U = 2195$, $p = 0.009$, both $p < 0.05$). 
This indicates that effective collaboration and information ease are key to task completion in this setting.

Regarding the qualitative data, one of the objectives of the Wizard-of-Oz technique was to make the participant believe that they are interacting with an automated agent and the qualitative feedback seemed to reflect this: \textit{``The AI in the game was not helpful at all [...]''} or \textit{``I was talking to Fred a bot assistant, I had no other partner in the game``}.

\paragraph{Single vs Multiple Wizards}
In Table \ref{tab:interaction_stats}, we compare various metrics from the dialogues collected with crowdsourcing with the dialogues previously collected in a lab environment for a similar task. Most figures are comparable, except the number of emergency assistant turns (and consequently the total number of turns). To further understand these differences, we have first grouped the dialogue acts in four different broader types: Updates, Actions, Interactions and Requests, and computed the relative frequency of each of these types in both data collections. In addition, Figures \ref{fig:dialogue_acts_amt} and \ref{fig:dialogue_acts_hri} show the distribution of the most frequent dialogue acts in the different settings. It is visible that in the lab setting where the interaction was face-to-face with a robot, the Wizard used more Interaction dialogue acts (Table \ref{tab:da_type_distribution}). These were often used in context where the Wizard needed to hold the turn while looking for the appropriate prompt or waiting for the robot to arrive at the specified goal in the environment. On the other hand, in the crowdsourced data collection utterances, the situation updates were a more common choice while the assistant was waiting for the robot to travel to the specified goal in the environment.

Perhaps not surprisingly, the data shows a medium strong positive correlation between task success and the number of Action type dialogue acts the Wizard performs, triggering events in the world leading to success ($R=0.475$). There is also a positive correlation between task success and  the number of Request dialogue acts requesting confirmation before actions ($R=0.421$), e.g., \textit{``Which robot do you want to send?''}. As Table 3 shows, these are relatively rare but perhaps reflect a level of collaboration needed to further the task to completion. Table \ref{tab:example_interaction} shows one of the dialogues collected where the Emergency Assistant continuously engaged with the Operator through these types of dialogue acts. 

The task success rate was also very different between the two set-ups. In experiments reported in \cite{HRILopes2019}, 96\% of the dialogues led to the extinction of the fire whereas in the crowdsourcing setting only 9.66\% achieved the same goal. In the crowdsourced setting, the robots were slower moving at realistic speeds unlike the lab setting\footnote{There was no live connection with the simulated physical environment implemented.}. A higher bonus and more time for the task might lead to a higher task success rate. 

\paragraph{Limitations}
It is important to consider the number of available 
participants ready and willing to perform the task at any one time. 
This type of crowdsourcing requires two participants to connect within a few minutes of each other to be partnered together. As mentioned above, there were some issues with participants not collaborating and these dialogues had to be discarded as they were not of use\footnote{Participants who collaborated still received the full payment regardless of their partner's behaviour.}. 

\subsection{Future Work}

In future work, we want to expand and improve the platform. Dialogue system development can greatly benefit from better ways of obtaining data for rich task-oriented domains such as ours.  Part of fully exploiting the potential of crowdsourcing services lies in having readily available tools that help in the generation and gathering of data. One such tool would be a method to take a set of rules, procedures or business processes and automatically convert to a FSM, in a similar way to \cite{lemon2008dude}, ready to be uploaded to the Wizard interface.

Regarding quality and coherence, dialogues are particularly challenging to automatically rate. In our data collection, there was not a correct or wrong dialogue option for the messages that the Emergency Assistant sent during the conversation, but some were better than others depending on the context with the Operator. This context is not easily measurable for complex tasks that depend on a dynamic world state. Therefore, we leave to future work automatically measuring dialogue quality through the use of context.

The introduction of Instructional Manipulation Checks \cite{Oppenheimer2009} before the game to filter out inattentive participants could improve the quality of the data (Crowdworkers are known for performing multiple tasks at once). \newcite{Goodman2013} also recommend including screening questions that check both attention and language comprehension for AMT participants. Here, there is a balance that needs to be investigated between experience and quality of crowdworkers and the need for large numbers of participants in order to be quickly paired.

We are currently exploring using the data collected to train dialogue models for the emergency response domain using Hybrid Code Networks \cite{williams-etal-2017-hybrid}.

\section{Conclusion}

In conclusion, this paper described a new, freely available tool to collect crowdsourced dialogues in rich task-oriented settings. By exploiting the advantages of both the Wizard-of-Oz technique and crowdsourcing services, we can effortlessly obtain dialogues for complex scenarios. The predefined dialogue options available to the Wizard intuitively guide the conversation and allow the domain to be deeply explored without the need for expert training. These predefined options also reinforce the feeling of a true Wizard-of-Oz experiment, where the participant who is not the Wizard thinks that they are interacting with a non-human agent. 



As the applications for task-based dialogue systems keep growing, we will see the need for systematic ways of generating dialogue corpora in varied, richer scenarios. This platform aims to be the first step towards the simplification of crowdsourcing data collections for task-oriented collaborative dialogues where the participants are working towards a shared common goal. The code for the platform and the data are also released with this publication.






\section{Acknowledgements}

This work was supported by the EPSRC funded ORCA Hub (EP/R026173/1, 2017-2021). Chiyah Garcia's PhD is funded under the EPSRC iCase EP/T517471/1 with Siemens.

\begin{table*}[ht]
\begin{center}
\begin{tabularx}{\textwidth}{l X l}
    \hline
    \textbf{Participant} & \textbf{Utterance} & \textbf{Dialogue Acts}\\
    \hline
    Assistant: & \textit{``Hi, my name is Fred, and I'm your emergency assistant''} & intro\_hello \\
     & \textit{``Are you there?''} & request\_attention \\
    Operator: & \textit{``Hi Fred, I am \_''} \\
    Assistant: & \textit{``Emergency alarm went off in processing module east tower. We have 4:56 to avoid evacuation''} & inform\_alert\_emergency \\
    Operator: & \textit{``Ok what do you suggest we do first''} \\
    Assistant: & \textit{``Do you want to make a PA announcement to evacuate processing module east tower?''} & request\_pa\_announcement \\
    Operator: & \textit{``Yes that sounds good''} \\
    Assistant: & \textit{``Action performed''} & action\_performed \\
     & \textit{``First, I'm activating emergency shutdown for processing module east tower''} & inform\_activate\_emergency\_shutdown \\
    Operator: & \textit{``Ok''} \\
    Assistant: & \textit{``What type of robot do you want to use to perform the inspection?''} & request\_robot\_type \\
    Operator: & \textit{``I would like to use the quad copter 1''} \\
    Assistant: & \textit{``Moving quad copter 1 to processing module east tower''} & inform\_moving \\
     & \textit{``The estimated time of arrival is 9 seconds''} inform\_robot\_eta \\
     & \textit{``There is still 3:08 before evacuation''} & inform\_time\_left \\
     & \textit{``Quad copter 1 has arrived to processing module east tower''} & inform\_arrival \\
    Operator: & \textit{``Is Quad copter indicating what the problem is?''} \\
    Assistant: & \textit{``Quad copter 1 is going to inspect processing module east tower''} & inform\_inspection \\
     & \textit{``The robot is reporting a major fire in the east tower gas compressor''} & inform\_emergency\_status \\
     & \textit{``Which robot should we send to put out the fire?''} & request\_robot\_emergency \\
    Operator: & \textit{``Should we extinguish the fire now using quad copter 2''} \\
    Assistant: & \textit{``Sending quad copter 2 to processing module east tower''} & inform\_moving \\
     & ... \\
    \hline
\end{tabularx}
\caption{Interaction between participants from one of the dialogues collected. 
}
\label{tab:example_interaction}
\end{center}
\end{table*}

\section{Bibliographical References}
\label{main:ref}
\bibliographystyle{lrec}
\bibliography{main}


\end{document}